\documentclass[showkeys,twocolumn,amsmath,amssymb]{revtex4}
\usepackage{hyperref}
\usepackage[dvips]{graphicx}
\usepackage{color}
\usepackage{amsfonts}
\usepackage{amsthm}
\usepackage{hyperref}
\usepackage{caption}
\usepackage{verbatim}
\usepackage{array}
\usepackage{dcolumn}
\usepackage{bm}
\newtheorem{defn}{Definition}[section]
\newtheorem{lem}[defn]{Lemma}
\newtheorem{theor}[defn]{Theorem}

\begin{document}
\title{Geometric definition of a new skeletonization concept.}

\author{Yannis Bakopoulos}
 \homepage{http://cag.dat.demokritos.gr}
 \email{ybak@dat.demokritos.gr}
\author{Theophanis Raptis}%
 \email{rtheo@dat.demokritos.gr}
\affiliation{
NCSR "Demokritos"\\
Computational Applications Group
}%

\author{Doxaras Ioannis}
\email{doxaras@inp.demokritos.gr}
\affiliation{
National Technical University of Athens\\
Department of Applied Mathematics \& Physics}

\date{\today}

\begin{abstract}
The Divider set, as an innovative alternative concept to maximal
disks, Voronoi sets and cut loci, is presented with a formal
definition based on topology and differential geometry. The relevant
mathematical theory by previous authors and a comparison with other
medial axis definitions is presented. Appropriate applications are
proposed and examined.
\end{abstract}


\maketitle

\tableofcontents 

\section{Introduction}\label{sec:1}
Starting from the pioneering paper of H. Blum~\cite{Blum}, the
medial axis as a descriptor and classifier of shapes and figures has
been established as the best defined and studied mathematical
concept in reference to thinning and skeletonization of contours and
shapes~\cite{Sherbrooke1,Sherbrooke2,Hyeong,Choi,Yang,Tsang1,Tsang2,
Sakellariou}. From the various mathematical tools (Maximal disks,
cut loci, Voronoi
sets~\cite{Sherbrooke1,Sherbrooke2,Hyeong,Choi,Yang,Tsang1,Tsang2,Blum,
Sakellariou}, the maximal disk method seems to be the most well
studied and applied, both in mathematical definition and
properties~\cite{Sherbrooke1,Hyeong,Yang,Tsang1} and in
applications~\cite{Sherbrooke2,Choi,Tsang2,Sakellariou}. Its
definition is best presented in the following form:

\begin{defn} Let $D$ be a subset of $\Re^n$. A closed ball $B$ (or disk in 2-D)
is said to be \emph{maximal} in $D$ if it is contained in $D$ and if
$B \subset B'$, where $B'$ is another closed ball, also contained in
$D$, then $B = B'$~\cite{Sherbrooke1,Hyeong}.
\end{defn}

The notion of maximal disks is based on the Euclidean metric.
\begin{equation}\label{eq:1}
d_E(\mathbf{x},\mathbf{y})=\sqrt{(x_1-x_2)^2 + (y_1 - y_2)^2}
\end{equation}

There are two other equivalent metrics known from any textbook of
real analysis:

\begin{equation}\label{eq:2}
d_{max}(\mathbf{x},\mathbf{y})=max\{|x_1-x_2|,|y_1-y_2|\}
\end{equation}

Called here the maximum coordinate metric. And the addition metric:
\begin{equation}\label{eq:3}
d_1(\mathbf{x},\mathbf{y})=\{|x_1-x_2|+|y_1-y_2|\}
\end{equation}

The above definition cannot be applied to one of the above given
metrics. The results would not lead to a proper skeleton. In most
cases it would lead to disconnected medial axes, contrary to the
definition of a skeleton. On the other hand, the Euclidean matrix
has serious problems when applied in discrete bitmap images [9].
Furthermore, due to its use of a square root function, it is more
calculation intensive than the other metrics.

In a completely different field and with entirely different
motivations, some similar concepts were developed by \emph{P. C.
Stavrinos} in the early 80's~\cite{Stavrinos,Stavrinos1}. The author
was attempting to develop better tools for the classification and
extraction of features of various geometric constructions, such as
classes of two dimensional manifolds immersed in a three dimensional
Euclidean space. The tools would be developed for applications in
various branches of mathematics and physics, for example in knot
theory, convexity, flows, the study of differential equations and
the propagation of their solutions and corresponding singularities,
as described by the Huygens principles~\cite{Ruuth,Fox}.

The developed concepts described global characteristics of surfaces
connecting them to local geometric and topological features. The
concepts presented in~\cite{Stavrinos,Stavrinos1}, mainly what the
author called \emph{the first curvature of locally convex type},
were extended and generalized in~\cite{Bakopoulos1,Bakopoulos2} and
some new concepts, presented here in Section 2, were introduced
there for the first time. At that time, the authors
of~\cite{Bakopoulos1} and~\cite{Bakopoulos2} were largely unaware of
the potential of their work in morphological applications. Their
interest was mainly in developing new mathematical concepts of a
\emph{not quite local but not fully global} nature, as tools in the
study and classification of geometric objects and properties. Some
such results on non--analytic curves embedded in $\Re^n$, convexity
and knots' theory will be published elsewhere.

When the research was turned towards morphology, the concept of the
contact disk and the Divider set, as an alternative skeleton method,
were presented in~\cite{Aggarwal}. The concepts were presented in a
discrete lattice environment and the maximum coordinate metric was
used.

In this work, a full definition of the concepts of the contact
curvature, the contact circle and the Divider set will be presented.
Emphasis will be put on the Euclidean metric method, on the
Euclidean plane and three dimensional space, although, in the
discussion sector, a comparison with the discrete version will be
attempted, with reference to specific applications, such as OCR and
robotic navigation. In Section \ref{sec:2}, a brief review of
previous ideas and definitions will be given, based mainly
on~\cite{Bakopoulos1}. In Section \ref{sec:3}, the basic definitions
will be presented, in the form of a set of non linear algebraic
equations and inequalities. Some fundamental results will be
presented. Discussion and conclusions will be the subject of Section
\ref{sec:4}.

\section{The curvature of locally convex type}\label{sec:2}
In previous work~\cite{Stavrinos,Stavrinos1}, the following
definitions were given:

\begin{defn}
A surface $S$ in the Euclidean space $\Re^3$ will be of locally
convex type iff, $\forall\,p\in S$, given $0<r_0$ , and $r\in
(0,r_0)$ the intersection $B^3(p,r)\bigcap S$, where $B^3(p,r)$ a
closed ball with center $p$ and radius $r$, will be the closure of a
simply connected coordinate neighborhood of $p$ on $S$.
\end{defn}

\begin{defn}
$\forall p\in S$, the curvature of locally convex type,
$K_{l.c.t}^{(1)}$, will be defined as the inverse of the infimum
$r_{l.c.t.}$ of all radii of closed balls $B^3(p,r)$ for which the
intersection $B^3(p,r)\bigcap S$ will be disconnected\footnote{If
the set of such radii r for which the above relation holds is empty,
which means that there are no closed balls such that their
intersection with S will be disconnected, then $K_{l.c.t}^{(1)}=0$,
as the inverse of the infimum of an empty set. The simplest examples
of such a situation are plane or spherical surfaces.}.
\end{defn}

In the above, $S$ was a connected, piece by piece smooth surface,
which means that it was the union of a finite set once continuously
differentiable surfaces~\cite{Stavrinos,Stavrinos1}.

In~\cite{Bakopoulos1}, an extension of the curvature of locally
convex type was given. The definition now included any point $p\in
\Re^3$:

\begin{defn}[a]
$\forall p\in \Re^3$, the curvature of locally convex type,
$K_{l.c.t}$, will be defined as the inverse of the infimum
$r_{l.c.t.}$ of all radii $r$ of closed balls $B^{(3)}(p,r)$ for
which the intersection $B^{(3)}(p,r)\bigcap S$ will be disconnected.

\begin{equation*}
r_{l.c.t}=\text{infim}\{r/\, 0<r \rightarrow B^{(3)}(p,r) \bigcap S
\text{ disconnected}\}.
\end{equation*}

\end{defn}

Another generalization of Def. 2.1 is now possible:

\begin{defn}[b]
A surface $S$ in $\Re^3$ will be of locally convex type if there is
no sequence of points $p_i\in \Re^3$, $i\in\aleph$, the set of
natural numbers, converging to a limit point $p\in \Re^3$, where
$0<K(p)<\infty$, such that:

\begin{equation}
\lim_{p_i\rightarrow p}{K(p_i)}<\infty
\end{equation}

This definition should be compared with assumptions about the
suitability of analytic curves, as stated in~\cite{Hyeong}. Whether
this definition is equivalent to analyticity, is for now an open
question for the authors of this work.
\footnote{In~\cite{Bakopoulos1}, the definition is given for
manifolds $M^d$ in $\Re^n$. It can be applied equally well to curves
$S$ in $\Re^2$ or $\Re^3$, as well as surfaces in $\Re^3$.}
\end{defn}

The next result has to do with the set of all points in $\Re^3$ for
which $0<K_{l.c.t}$: This set has been named $\Pi(S)$.

\begin{theor}
The set $\Pi(S)$ of a curve $S$ in $\Re^2$ or $\Re^3$, or a surface
$S$ in $\Re^3$ is open by the Euclidean topology\footnote{For a
detailed proof, see~\cite{Bakopoulos1}.}.
\end{theor}

In~\cite{Bakopoulos1}, a series of results follows, concerning
necessary and sufficient conditions for a point $p\in \Re^2$ or
$\Re^3$ to belong to $\Pi(S)$, $S$ being a curve or a surface. The
most important and utile results will be presented here, without
proof, as follows (For detailed proofs, see~\cite{Bakopoulos1}):

\begin{lem}
If and only if a curve $S$ is \emph{(a)} a circle $S^1$, or
\emph{(b)} a connected one dimensional subset of a straight line,
the set $\Pi(S)$ will be empty.
\end{lem}

\begin{lem}\label{lem:1}
If $S$ is a curve in $\Re^2$ and is homeomorphic to $(0,1),\,(0,1]$
or $[0,1]$, then a necessary and sufficient condition for a point
$p\in \Re^2$ to belong to $\Pi(S)$ is that a compact, connected,
proper subset $Q$ of $S$ exists, which is a circular arc with $p$ as
its center and disconnects $S$ by not containing any endpoints.
Therefore the relation:

\begin{equation*}
d(p,q)=\text{constant holds } \forall\, q \in\, Q\{d(p,q)\}
\end{equation*}

Furthermore, there exists an open neighborhood $N(Q)$ of $Q$ such
that $\forall q\in Q$ and $q'\in (N(Q)-Q)$, it is true that
$d(p,q')<d(p,q)$.
\end{lem}

The compact subset $Q$ may be a single point, which is the case if
$S$ is analytic and contains no constant curvature arcs, as
postulated in~\cite{Hyeong}, for example.

\begin{lem}
If $S$ is a curve in $\Re^2$ and is homeomorphic to $S^{(1)}$, then
a necessary and sufficient condition for any point $p\in \Re^2$ to
belong to $\Pi(S)$ is either the existence of two points $q_1$,
$q_2$, and corresponding open neighborhoods $N_1(q_1)$ and
$N_2(q_2)$ such that: $\forall q_i'\in (N_i(q_i)-Q_i)$, the
relation: $d(p,q_i')<d(p,q)$ will hold, $i=1,2$, or the existence of
two compact, connected, proper subsets $Q_i$, with corresponding
open neighborhoods $N_i(Q_i)$ will exist, which will be circular
arcs having $p$ as a center and the relation of Lemma \ref{lem:1}
holds: $\forall q_i\in Q_i$ and $\forall \,q_i'\in (N(Q_i)-Q_i)$, it
is true that $d(p,\,q_i')<d(p,\,q_i)$, $i=1,\,2$.
\end{lem}

The conclusion of~\cite{Bakopoulos1} (\emph{Theorem 3.11, p. 279}),
was that, for the calculation of the curvature of locally convex
type, $K_{l.c.t}$ of a curve $S$, at any point $p\in \Re^2$, all
perpendiculars from $p$ to $S$ should be found and those for which
the following property holds, should be selected:
$d(p,q_i')>d(p,q_i)$, where $q_i$, $q_i'$ as in the previous lemma,
$i=1,2,\ldots,n$, if $n$ such perpendiculars exist. Then the
distances should be put in increasing order, $d(p,q_1)\leq
d(p,q_2)\leq \ldots$. Then, the curvature of locally convex type
would be the inverse of the second distance in the ascending order
series:

\begin{equation*}
K_{l.c.t}(p)=\frac{1}{d(p,q_2)}
\end{equation*}

A simple corollary of the above results is the close correspondence
of $\Pi(S)$ and the curvature of locally convex type to the evolute
of a curve. For example, the ellipse has as $\Pi(S)$ the area
enclosed by its evolute~\ref{fig:1}. A similar result is true of the
parabola~\ref{fig:2}. Specifically, the set $\Pi(S)$, if $S$ is a
parabola, is the area bounded by the convex side of the evolute. As
it will be demonstrated in the next Section \ref{sec:3}, the
symmetries of a curve, the symmetries of its evolute and some other
geometric characteristics will decide the topological properties of
the Divider. A very characteristic case is the hypotrochoid and its
evolute~\ref{fig:3}. In this case, the symmetry of the curve and its
evolute and most of all, the multiplicity of branches of the evolute
which turn their convex side to a certain area will decide the
multiplicity of branches of the Divider radiating from the center of
the hypotrochoid\footnote{Proof of all above results
in~\cite{Bakopoulos1}.}~\ref{fig:3}.

The above results were applied to the derivation of the concepts
presented in the next Section \ref{sec:3}. These are the contact
directions, the contact circles, the contact curvature and the
Divider set.

\section[Contact disks \& the Divider.]{The contact disks and the Divider set of a curve or surface}\label{sec:3}
A series if definitions is in order. They are based on the previous
concepts and are a natural continuation of the previous theory.

\begin{defn}
Let a curve $S$ in $\Re^2$ be once continuously differentiable and
have a perpendicular at each point. Let $p\in S$ and let $B(k,r)$ be
a closed disk of center $k$ and radius $r=d(k,p)$. Furthermore, let
$S\bigcap B(k,r)=\{p\}$. Then the direction $|p,k|$ will be called a
contact direction of $S$ in $p$.
\end{defn}

A similar concept can be defined for a curve or a surface $S$ in
$\Re^3$. Then the two dimensional disk is replaced by a three
dimensional ball.

The fact is that in every case the point $p$ is the only common
point of $S$ and $B(k,r)$. If $p$ is not an end point and given the
assumptions of a continuous first derivative and an unambiguously
defined perpendicular, there are two contact directions at each $p$
in a curve in $\Re^2$ or a surface $S$ in $\Re^3$. They are the two
directions of the perpendicular. If, on the other hand, $S$ is a
curve in $\Re^3$, then there is a plane normal to the curve and
every direction on this plane is a contact direction.

But there are extreme cases, such as end points or singularities. In
those cases, there will be a whole cone of contact directions at
$p$. If $p$ is the end point of a plane curve, there is a semicircle
of directions with its diameter on the perpendicular and at the side
of it pointing away from the curve. If the curve or surface lies on
$\Re^3$, then there is a hemisphere of contact directions. In
general, the space of contact directions depends on the morphology
of the curve or surface $S$ in the neighborhood of $p$. The same
goes for singularities, where there is not a uniquely defined
tangent and perpendicular at $p$. If at a point $p$ on $S$ a convex
angle exists on one side and, naturally, a concave angle exists on
the other side of $S$, then the contact disk on the convex side will
have infinite curvature and the contact directions on the other side
will cover an area, disk section in $\Re^2$ or ball section in
$\Re^3$, whose shape will depend on the morphology of the
singularity (See the definitions of sharp and dull corner points
in~\cite{Hyeong}).

In each and every one of the above cases, the set of all disks or
balls, having a single contact point $p$ with $S$, is well defined.
If the disk is enlarged, by moving its center $k$ away from $p$ and
thereby enlarging its' radius $|p,q|$, one of two things will
happen. Either, at some point, the intersection $S\bigcap B(k,r)$
will contain other points of $S$, besides $p$, or the intersection
will contain only p all the way until $k$ reaches infinity. In the
former case, there must be a \emph{supremum} of the radii of such
disks having the property $S\bigcap B(k,r)=p$. In the second case
the \emph{supremum} is infinite. Then we have the following
definition.

\begin{defn}
The inverse of the radius of the \emph{supremum} of all disks which
have the property:

\begin{equation*}
S\bigcap B(k,r)=\{p\}
\end{equation*}

is called the \textbf{contact curvature} of $S$ at $p$. The locus of
all centers of the \emph{supremum} disks, $\forall p\in S$, is
called the \textbf{Divider} set of $S$.
\end{defn}

\begin{figure*}\label{fig:3}
\includegraphics[scale=.8]{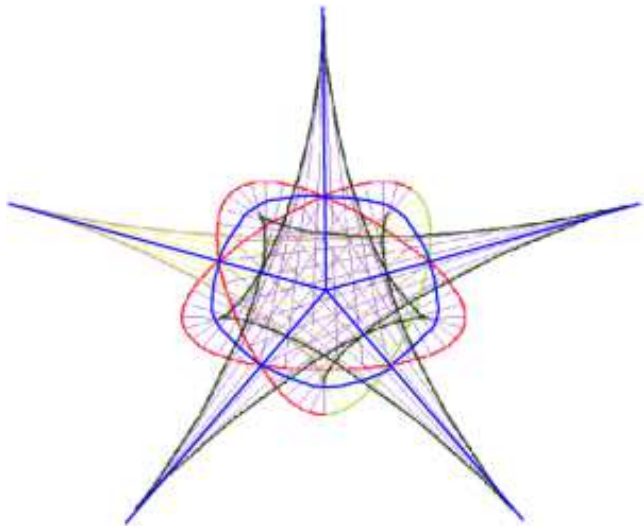}
\caption[Hypotrochoid Divider]{ A hypotrochoid $S$ and its evolute
are the best example of an exception to the rules for curves without
double points. Note that vertices of a curve correspond to its
evolute cusps. The lines are radii of tangent circles. The area
inside the evolute defines the set $\Pi(S)$ of the hypotrochoid $S$.
The points at the interior have a positive curvature of locally
convex type. The Divider interior points are contained in the set
$\Pi(S)$. The cusps of the evolute corresponding to the maximum
curvature points of the hypotrochoid, \textbf{are not end points of
the Divider}. This is because the osculating disks at the maximum
curvature points are larger than the corresponding maximal disks in
the same points. The cross points of the hypotrochoid are points of
the Divider with zero radius maximal disks, having infinite
curvature of locally convex type. The interior of the smaller curved
asteroid decagon inside the hypotrochoid is the area where the disks
defining the curvature of locally convex type may have an
intersection with the hypotrochoid with five distinct connected
components. As a special case, the central junction point of the
Divider, situated at the center of the Figure, is the starting point
of five branches of the Divider, these being straight lines
radiating from the junction to each cross point of the hypotrochoid.
There are five more branches of the Divider, uniting the
hypotrochoid cross points within each outer lobe of the
hypotrochoid. The last five branches of the Divider are straight
lines starting from each cross point of the hypotrochoid, passing
through the corresponding cusp of the evolute which is a center of a
minimum curvature osculating disk and going on to infinity. In this
case the Divider is a multiply connected graph containing singular
points of zero contact curvature due to the existence of analogous
singularities (cross points) in the hypotrochoid.}
\end{figure*}

It is obvious that the set of all disks with the property $S\bigcap
B(k,r)=\{p\}$ is linearly ordered by inclusion. This means that, as
the radius of each disk increases, it contains all disks with
smaller radii. It is equally obvious that, if $S$ is a closed curve,
in the sense of being homeomorphic to $S^{(1)}$, containing an area,
and the contact direction is taken towards the inside of the area,
then all disks with the above property are contained in the area
surrounded by $S$. Then, assuming that the \emph{supremum} is not a
maximum, the above definition is identical with that of a maximal
disk~\cite{Sherbrooke1,Sherbrooke2,Hyeong,Choi,Yang,Blum}. Also
obviously, since the contact direction is defined for both sides of
a closed contour, the Divider is equally well defined for an arc
homeomorphic to  a one dimensional connected subset of a straight
line or a one dimensional circle $S^{(1)}$, it may cover cases where
the maximal disk definition cannot be applied easily, or not at all.
Furthermore, since the above definitions may easily be extended to
curves with multiple cut points,~\ref{fig:3}, or even families of
disconnected curves or surfaces, the Divider concept can cover all
relative medial axis definitions, such as the Voronoi set, etc. This
advantage is expected to be decisive in the case of the
morphological study of a whole page of handwritten text, as a first
step for handwriting OCR.

By far the most important advantage of the Divider definition is the
fact that it can be easily and naturally extended in a discrete
lattice environment, by the use of another metric than the Euclidean
one~\cite{Aggarwal}. The rules for the construction of the contact
disks and the Divider in~\cite{Aggarwal} are a first tentative
attempt for the development of algorithms suitable for the discrete
lattice case. The advantages and potential applications of the
discrete lattice~\cite{Tsang1,Tsang2}, will be included in these
authors' future work. As already mentioned in~\cite{Aggarwal}, the
applications considered so far are OCR on handwritten text in a
discrete lattice environment and robotic navigation and map making
in complex enclosed spaces where outside navigational aids are not
available.

The results presented in Section \ref{sec:2} and the above
definitions lead to the simple conclusion that the Divider set is of
a curve S in the plane is included in the closure of $\Pi(S)$. Proof
of the above statement is provided if a series of well known results
from differential geometry are given here in the form of lemmas,
without proof.

\begin{lem}
If the evolute $E$ of a thrice continuously differentiable curve $S$
in $\Re^2$ is considered, the cusps of the evolute correspond to
points of minimum or maximum curvature of the osculating circles of
$S$ at those points. The cusps corresponding to maximum curvature
have the property that if a small open disk $B(k,\varepsilon)$ is
taken with the cusp as a center, then the cusp is closest to $S$
than any other point of the evolute within the disk. If an analogous
disk is taken at a cusp corresponding to minimum curvature, then the
cusp is furthest from $S$ than any other point of the evolute within
the disk.
\end{lem}

\begin{lem}\label{lem:2}
Let a curve $S$ in $\Re^2$ and its evolute $E$ be considered. By the
theory developed in~\cite{Bakopoulos1}, $\Pi(S)$ is the area in
$\Re^2$ in the convex side of the evolute. If $S$ is homeomorphic to
$S^{(1)}$, the area of $\Pi(S)$ will be enclosed within the evolute.
If, on the other hand, $S$ is a curve in $\Re^2$ and is homeomorphic
to $(0, 1)$, $(0, 1]$ or $[0, 1]$, then $\Pi(S)$ is the area of
$\Re^2$ toward which the evolute $E$ turns its convex side.
\end{lem}

\begin{figure}[!htp]\label{fig:1}
\centering 
\includegraphics[scale=1]{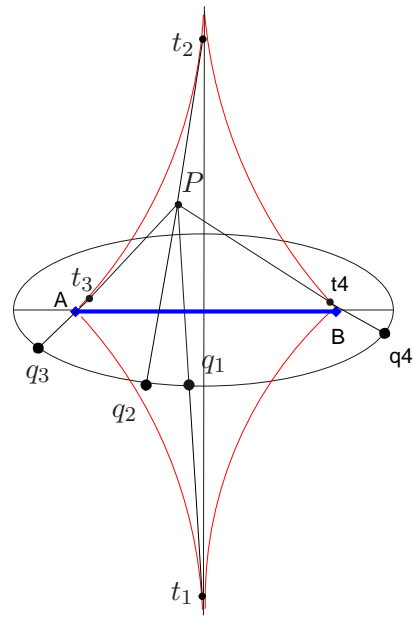}
\caption[Elipse Divider]{The set $\Pi(S)$ of an ellipse $S$ is the
set of points within the interior of the area within the evolute,
called, in this case an asteroid. It is indicated by the red line.
It is the set of all points in $\Re^2$ for which the curvature of
locally convex type takes a non zero value. The Divider of $S$ is
the segment of the great axis within the closure of $\Pi(S)$. Its
endpoints are the centers of osculating circles of maximum
curvature.}
\end{figure}

A typical example is an ellipse~\ref{fig:1}.

\begin{figure}[!htp]\label{fig:2}
\centering
\includegraphics{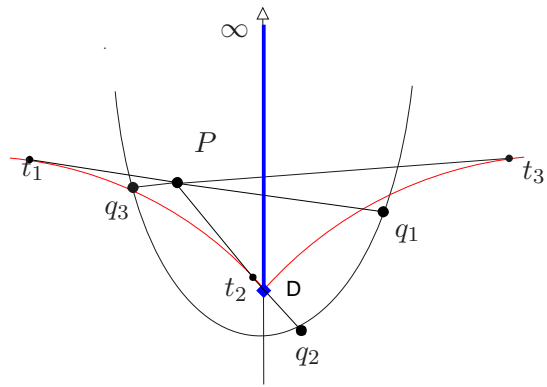}
\caption[Parabola Divider]{The evolute of a parabola is called a
Neile's parabola. The set $\Pi(S)$ lies within the area delimited by
the red line, where the evolute turns its convex side. The Divider
of a parabola is the part of its axis starting from the center of
the osculating circle at the vertex of maximum curvature and going
to infinity.}
\end{figure}

If, on the other hand, $S$ is a curve in $\Re^2$ and is homeomorphic
to $(0,\,1)$, $(0,\,1]$ or $[0,\,1]$, then $\Pi(S)$ is the area of
$\Re^2$ toward which the evolute $E$ turns its convex side. A
typical example is a parabola~\ref{fig:2}.

\begin{proof}
By the theory developed in~\cite{Bakopoulos1}, if $S$ is
homeomorphic to a one - dimensional connected subset of a straight
line, then a necessary and sufficient condition for the relation
$p\in \Pi(S)$ to be valid, is that there is at least one line $p$
and $q$ normal to $S$, such that the distance $|p,q|$ of $p$ from
$q$ is a local maximum of the distances of $p$ from points of $S$.
If, on the other hand, $S$ is homeomorphic to $S^{(1)}$, then two
line segments $|p,q_1|$, $|p,q_2|$, which must be local maxima of
the distance of $p$ from points of $S$~\cite{Bakopoulos1}. For the
above to hold, there must exist analogous tangent lines from $p$ to
the evolute $E$ of $S$. For the distance of $p$ from a point $q$ of
$S$ to be a local maximum, the point of contact $t$ of $pq$ with $E$
must lie between $p$ and $q$. The above statement can be true only
if $p$ is at the convex side of $E$, so that the appropriate tangent
or tangents may exist.
\end{proof}

\begin{lem}
If $S$ is homeomorphic to $S^{(1)}$ and different from $S^{(1)}$,
then from every point $p$ of $\Pi(S)$, there will be at least four
tangents to $E$ from $p$. Correspondingly, they will be normal to
$S$, by definition of involutes and evolutes. Two of them will
define local minimum distances of $p$ from $S$ and two will define
local maximum distances from $p$ to $S$~\ref{fig:1}. Each one of
then is tangent to $E$ at a point $t_i$, $i=1,2,3,4$, which is the
center of an osculating circle of $S$ at a point $q\in S$. The lines
defining local maxima and minima of distance will alternate in
succession as they meet $S$. If, on the other hand, $S$ is a curve
in $\Re^2$ and is homeomorphic to $(0, 1)$, $(0, 1]$ or $[0, 1]$ but
is not a straight line subset, then there will be at least three
tangents from $p$ to $E$. Two of them will define local minimum
distances of $p$ from $S$ and one of them will define a local
maximum distance. The line defining a local maximum will lie between
the other two lines. \footnote{By definition, every tangent to the
evolute $E$ will be normal to $S$ at $q$ and the osculating
curvature of S at $q$ will be $k(q)=1/|t,q|$. A theorem, including
results well known from the theory of differential geometry of
curves and surfaces, will be presented below. Similar results are
presented and proved in the relative literature (For example,
see~\cite{Sherbrooke1,Hyeong}).}
\end{lem}

\begin{theor}\label{theor:1}
If a tangent is defined from $p\in \Pi(S)$ to the evolute $E$ of
$S$, being normal to $S$ at $q$, then the points $p'$ on the
straight line $(p,q)$ have the following properties: If a point $p'$
lies on the far side of $q$ relative to the center $t$ of the
osculating circle, the distance $|p,q|$ is a local minimum, in the
sense that if a small enough open neighborhood of $q,N(q)$, is
considered on $S$, then $\forall q'\in \Pi(S)$ the relation:
$|p,q|<|p,q'|$ holds.
\end{theor}

The same is true if $p$ lies between $q$ and $t$. If $p$ is at the
far side of $t$ relative to $q$, in other words if $t$ is between
$p$ and $q$ then the distance $|p,q|$ is a local maximum, meaning
that, if a small enough open neighborhood of $q,N(q)$, is considered
on $S$, then $\forall q'\in\Pi(S)$ the relation: $|p,q|>|p,q'|$
holds. Furthermore, if $p=t$, then if $t$ is the center of an
osculating circle with maximum curvature, as described above, the
relation $|t,q|<|t,q'|$ holds. If $t$ is the center of an osculating
circle of minimum curvature, the relation $|t,q|>|t,q'|$ holds.
Finally, if $t$ is the center of an osculating circle with neither
maximum nor minimum curvature, the distance $|t,q|$ is neither
maximum nor minimum. If any neighborhood $N(q)$ is considered on
$S$, however small, the points $q'$ in $N(q)$ for which the
osculating curvature is larger than that in $q$ will lie outside the
osculating circle at $q$, while the points $q'$ where the osculating
curvature is larger than that of $q$ will lie within the osculating
circle ay $q$. The above results are well known facts from
differential geometry. An osculating circle of a curve at a point
$p$ is in contact of at least the second degree with the curve at
$p$. If the curvature is locally maximum or minimum, then the
contact is at least of the third degree. If the curvature is locally
a maximum, then the curve lies outside the osculating circle at a
neighbourhood of $p$. If the curvature is a local minimum, then the
curve lies inside the osculating circle at a neighbourhood of $p$.
In ordinary cases, i.e., when the contact at the point in question
does not happen to be of an order higher than the second, the circle
of curvature will not merely touch the curve, but will also cross
it~\cite{Taylor}. These results can be naturally extended to curves
and surfaces in three dimensions~\cite{Sherbrooke1,Hyeong,Taylor}.

Now an important Theorem will be presented:

\begin{theor}
Let $S$ be a curve in $\Re^2$, homeomorphic to $(0,1)$, twice
continuously differentiable, unbounded, not containing arcs of
constant curvature and with only one point $q_0$ of maximum
curvature. The Divider of $S$ is contained in the closure of
$\Pi(S)$, at the part of $\Re^2$ where its evolute $E$ turns its
convex side. The end points of the Divider are cusps of the evolute
$E$ of $S$, corresponding to maximum curvature of the relative
osculating circles.
\end{theor}

\begin{proof}
If $S$ is as above, obviously the evolute $E$ will have two branches
joining at a cusp $t$ and having a common tangent there. The cusp
$t$ will be the center of an osculating circle of maximum osculating
curvature. If a point $p\neq t$ belongs to the Divider, there will
be exactly three straight line segments $|p,q_1|$, $|p,q_2|$,
$|p,q_3|$, normal to $S$ and tangent to $E$. One of them, $|p,q_2|$,
will define a local (and global) maximum distance of $p$ from $S$
and will lie among the other two. The other two distances, $|p,q_1|,
|p,q_3|$, will be equal by definition of the Divider. Also, being
equal to the radius of the \emph{supremum} of disks having contact
with $S$ at only one point, it will also be the infimum of the disks
having more than one contact points with $S$. The set $B^{(2)}(p,
|p,q_2|)\bigcap S$ will be disconnected, containing only two
distinct points $q_1$ and $q_3$. Therefore, $p\in \Pi(S)$ On the
other hand, let the cusp t of $E$ be considered. It will be the
center of the circle of maximum osculating curvature of $S$ at $q$.
As such, by Lemma \ref{lem:2} and \ref{theor:1}, it will also be the
center of the maximum disk contacting $S$ at its single point $q$.
Therefore, the disk having $t$ as a center and $|t,q|$ as a radius,
will define the contact curvature of $S$ at $q$. The cusp t will
belong to the Divider and since it will not belong to $\Pi(S)$, it
will be an end point of the Divider. The osculating disk will not
intersect $S$ at any other point, since $S$ has only one point of
maximum curvature. Similar results can be easily proved for more
general curves, homeomorphic to $(0,1]$, $[0,1]$, $S^{(1)}$, lines
which are receptive of a normal at each point but their osculating
curvature function is piece by piece continuous, or even
disconnected sets of curves. The above results do not hold only in
cases where S has points of self intersection~\ref{fig:3}.
\end{proof}

All the above statements can be verified by the Divider defining
equations presented below: Let a curve $S$ in $Â$ be defined by the
parametric functions:

\begin{eqnarray}
x_1 &=& x_1(t)\\
x_2 &=& x_2(t)\\ \nonumber
\end{eqnarray}

Given any value $t_1$ of the defining parameter of $S$, the
following quantities will be calculated, by the equations and
inequalities presented below: $x_1(t_1),\,x_2(t_1),\,t_2,$
$x_1(t_2),\,x_2(t_2),\,x_{10},\,x_{20}$.

\begin{eqnarray}\label{eq:4}
x_1&=&x_1(t_1)\\\label{eq:5} x_2&=&x_2(t_1)\\\label{eq:6}
x_1&=&x_1(t_2)\\\label{eq:7}
 x_2&=&x_2(t_2)\\ \nonumber
\end{eqnarray}

\begin{widetext}
\begin{eqnarray}\label{eq:8}
(x_1(t_1)-x_{10})\frac{dx_1(t_1)}{dt_1}+(x_2(t_1)-x_{20})\frac{dx_2(t_1)}{dt_1}&=&0\\\label{eq:9}
(x_1(t_2)-x_{10})\frac{dx_1(t_2)}{dt_2}+(x_2(t_2)-x_{20})\frac{dx_2(t_2)}{dt_2}&=&0
\\ \nonumber
\end{eqnarray}

\begin{eqnarray}\label{eq:10}
(x_1(t_1)-
x_{10})^2+\left(x_2(t_1)-x_{20}\right)^2&=&(x_1(t_2)-x_{10})^2+\left(x_2(t_2)-x_{20}\right)^2\\
\nonumber
\end{eqnarray}
\begin{eqnarray}\label{eq:11}
0<(x_1(t_1)-x_{10})\frac{d^2x_1(t_1)}{dt_1^2}+\left(\frac{dx_1(t_1)}{dt_1}\right)^2+
(x_2(t_1)-x_{20})\frac{d^2x_2(t_1)}{dt_1^2}+\left(\frac{dx_2(t_1)}{dt_1}\right)^2\\\label{eq:12}
0<(x_1(t_2)-x_{10})\frac{d^2x_1(t_2)}{dt_2^2}+\left(\frac{dx_1(t_2)}{dt_1}\right)^2
+(x_2(t_2)-x_{20})\frac{d^2x_2(t_2)}{dt_2^2}+\left(\frac{dx_2(t_2)}{dt_2}\right)^2\\
\nonumber
\end{eqnarray}

There are important cases where the equations:
\begin{eqnarray}\label{eq:13}
(x_1(t_1)-x_{10})\frac{d^2x_1(t_1)}{dt_1^2}+\left(\frac{dx_1(t_1)}{dt_1}\right)^2
+(x_2(t_1)-x_{20})\frac{d^2x_2(t_1)}{dt_1^2}+\left(\frac{dx_2(t_1)}{dt_1}\right)^2=0
\\\label{eq:14}
(x_1(t_2)-x_{10})\frac{d^2x_1(t_2)}{dt_2^2}+\left(\frac{dx_1(t_2)}{dt_1}\right)^2
+(x_2(t_2)-x_{20})\frac{d^2x_2(t_2)}{dt_2^2}+\left(\frac{dx_2(t_2)}{dt_2}\right)^2=0\\
\nonumber
\end{eqnarray}
\end{widetext}

are true and yet the distances $|p,q_1|$, $|p,q_2|$ are local
minima. In that case, the sign and order of the lowest, in order
succession, nonzero derivative decides if $|p,q_1|$ and $|p,q_2|$
are local maxima or local minima of the distances of p from the
points $q$ of $S$\footnote{Relative theorems are contained in most
reference books of infinitesimal calculus}. This simplified version
of minimum distance conditions is given here for reasons of space.
In cases of curves or surfaces with boundaries, the existence of
local maxima or minima of the distance of $p$ from $S$ is not
necessarily connected with the existence of normals from $p$ to $S$.
In such cases, minimum or maximum distances may lie on the boundary
of $S$, in which case a set of relations analogous to \ref{eq:1},
\ref{eq:2}--\ref{eq:3}, \ref{eq:4} may not exist, or may yield only
partial results, in the creation of the Divider. In such cases, the
algorithms for the creation of the Divider should use procedures not
entirely based on such well defined and solvable equations.

As $t_1$ spans $S$, the coordinates $x_{10}$, $x_{20}$, of the
Divider are calculated, as functions of the variable $t_1$.

Let a curve $S$ in $\Re^2$, homeomorphic but not similar to $S(1)$,
be considered. The above equations are consistent with the
definition of maximal
disks~\cite{Sherbrooke1,Sherbrooke2,Hyeong,Choi,Yang,Blum}. The
maximal disks definition and the Divider definition are equivalent
as long as the Euclidean metric \ref{eq:1} is used. The definition
of contact disks and maximal disks as stated in this work and
in~\cite{Aggarwal} are not equivalent to the traditional definition
of maximal disks, if one of the other metrics, \ref{eq:2} and
\ref{eq:3}, is used.

The above equations \ref{eq:4} to \ref{eq:10} are designed to find
the coordinates $x_{10}$,  of all points $x_1(t_1)$ and $x_2(t_1)$
which belong to the Divider of $S$. The procedure is as follows. If
a value $t_1$ of the parameter $t$ in the relations \ref{eq:4} and
\ref{eq:5} defining $S$ is given, then, by \ref{eq:4} and
\ref{eq:5}, a point $q_1$ with coordinates $x_1(t_1)$ and $x_2(t_1)$
will be defined. Equations \ref{eq:6} and \ref{eq:7} define a new
value $t_2$ of $t$ and the coordinates $x_1(t_2)$ and $x_2(t_2)$ of
a corresponding point $q_2$, with appropriate properties designated
by the following equations. The rest of the equations, \ref{eq:9} to
\ref{eq:10}, define the coordinates $x_{01}$ and $x_{02}$ of a point
$p\in \Re^2$ which, as stated above, will belong to the Divider of
S. Equation
 \ref{eq:8} signifies that $|p,q_1|$ is normal to $S$. Equation
 \ref{eq:9} signifies that $|p,q_2|$ is normal to $S$. Equation
 \ref{eq:10} signifies that $|p,q_1| = |p,q_2|$. Finally,
inequalities \ref{eq:11}, \ref{eq:12} certify that $|p,q_1|$ and
$|p,q_2|$ are local minima of the distances of $p$ from the points
of $S$.

The above equations can be easily extended to curves and surfaces in
$\Re^3$. Other, more general geometric objects in abstract metric
spaces may be defined by similar equations. By the theory of curves
and surfaces, if $S$, be it a curve or a surface in $\Re^2$ or
$\Re^3$, is unbounded in the topological sense and, furthermore, has
no boundary points by its intrinsic topology~\cite{Spivak}, the set
of equations \ref{eq:4} to \ref{eq:10} has at least one solution. In
$\Re^2$, this would mean a minimum of distances $|p,q|$, $q\in S$,
if one of the inequalities \ref{eq:11} or \ref{eq:12} would hold.
Conversely, if none of the above inequalities holds and the solution
yields a local maximum, there will be at least two more solutions,
both defining local minima of distances $|p,q|$, $q\in S$, at points
$q_1$, $q_2$. By \ref{eq:10}, $|p,q_1| = |p,q_2|$. Then the point
$p$ would belong to the Divider of $S$. Analogous results hold for
bounded curves or surfaces, homeomorphic to $S(1)$ or $S(2)$ but not
having everywhere constant curvature. A special case is that of a
center $c$ of an osculating circle of constant curvature at every
point $q'$ on $N(q)$. In that case the following are true:

\begin{enumerate}
\item S contains a compact, connected circular arc $Q$ with $c$ as its
center.
\item Every contact direction for every $q\in Q$ passes through c.
\item The disk of the osculating circle is the supremum of the disks
which have with $S$ only one common point $q$, belonging to $Q$. The
center $c$ of the osculating circle for the points $q$ of $Q$ is the
corresponding point for the Divider for $Q$.
\end{enumerate}

All the above results are presented here without proof, since they
are well known from classical differential geometry\footnote{See
also pertinent results in~\cite{Sherbrooke1,Sherbrooke2}}.

\section{Discussion and conclusions}\label{sec:4}
In the above, the main contribution is the definition of a new
skeleton concept, the Divider set. Being in some sense the
\emph{reverse} of the maximal disk definition, the definition of the
Divider is given in reference to the points of a curve or surface.
This may or may not be the boundary of an area, finite or infinite.
The definition and the construction of a maximal contact disk will
start from a point on the curve or surface. On the other hand, the
classical definition of Blum, has to do with maximal inscribed
disks. Therefore, Therefore the algorithms creating it will be in
principle more efficient from algorithms referring to all points
inside a given enclosed area, as are the grassfire or wave front
algorithms and most maximal disks algorithms these authors are aware
of~\cite{Sherbrooke1,Sherbrooke2,Hyeong,Choi,Yang,Tsang1,Tsang2,Blum,
Sakellariou}. This seems to be the case in the first attempts for a
Divider creating algorithm by the use of the defining equations
using the Euclidean metric \ref{fig:1}, \ref{fig:2}, \ref{fig:3},
and also in the case of the discrete lattice~\cite{Aggarwal}. In
that last case, the maximum coordinate metric is used. In most cases
in the literature, for a small sample
see~\cite{Tsang1,Tsang2,Sakellariou}, the Euclidean metric is used
for the creation of skeletons, even in a discrete lattice
environment.

The maximal disk definition, as mentioned above, is not well suited
for the maximum coordinate metric. On the other hand the Divider
definition can be naturally modified~\cite{Aggarwal} for a discrete
lattice environment. The result is a connected set of cells, having
at most a two cells width wherever the distance of the lines is an
even number of cells. This is a well known problem with discrete
lattices~\cite{Tsang1,Tsang2,Sakellariou} and there are many issues
to be discussed. One of them, arguably the most important, is the
creation of a skeleton suitable for image compression and more or
less faithful restoration~\cite{Sherbrooke1,Hyeong,Choi,Yang}. The
authors of this work are of the opinion that in some cases
reproduction of a faithful image requires that all skeleton pixels
should be kept and no further thinning algorithm should be applied.
In other cases, like OCR, where no faithful reconstruction of the
image is required but the objective is the extraction and
preservation of some important features of the initial image, there
are some simple thinning algorithms that can be applied with good
results~\cite{Aggarwal}.

In general, the Divider seems to work in a satisfactory manner for
the applications of interest to the authors, for example OCR and
robotic navigation in enclosed environments, among others. To
conclude, by the work so far, the Divider concept seems to have some
distinct advantages compared with many other \emph{state of the art}
skeletonization methods. It has a precise mathematical definition,
easily implemented algorithms for two and three dimensional
Euclidean spaces utilizing the Euclidean metric, as well as discrete
lattice spaces, utilizing the maximum coordinate metric. It seems to
be promising for many applications, although further testing and
comparison with alternate methods is still to be done in future
work. It has especially promising attributes in specific interesting
applications, such as handwritten text recognition and robotic
navigation, among others.

\begin{acknowledgements}
The authors would like to thank NCSR "Demokritos" for warm
hospitality and Pr. P.C. Stavrinos for fruitful discussions.
\end{acknowledgements}

\end{document}